\documentclass[cup6a]{cupbook}
\usepackage{amssymb,graphicx,natbib}

\begin{document}
\pagenumbering{arabic}

\chapter[M.G.H.Krause \& V.~Gaibler: Jet related emission line regions]{
  Physics and fate of jet related emission line regions\\
  Martin~G.~H.~Krause\\
  Max-Planck-Institut f\"ur extraterrestrische Physik, Giessenbachstraße,
  85748 Garching, Germany and\\
  Universit\"atssternwarte M\"unchen, Ludwig-Maximilians-Universit\"at, 
  Scheinerstr. 1, 81679 M\"unchen, Germany\\
Volker Gaibler\\
  Max-Planck-Institut f\"ur extraterrestrische Physik, Giessenbachstraße,
  85748 Garching, Germany\\
 }

\vspace*{2cm}


\section{Introduction} At redshifts above $z\gtrsim0.5$ extragalactic
jet sources are commonly associated with extended emission line
regions \citep[for a review see][]{McC93,MdB08}. The most prominent
emission line is the hydrogen Lyman~$\alpha$ line, but other typical
nebular emission lines have also been found.  These regions are up to
100~kpc in extent, anisotropic and preferentially aligned with the
radio jets (\textit{alignment effect}).  Their properties correlate
with the ones of the radio jets: Smaller radio jets ($<100$~kpc) have
more extended emission line regions with larger velocity widths
(1000~km~s$^{-1}$) that are predominantly shock ionised, as diagnosed
from their emission line ratios. Larger radio jets ($>100$~kpc) have
emission line regions even smaller than 100~kpc. Their turbulent
velocities are typically about 500~km/s and the dominant excitation
mechanism is photo-ionisation.

The physical function of these emission line regions can be compared
to a detector in a particle physics experiment: In both cases a beam
of high energy particles hits a target. Analysis of the interactions
in the surrounding detector, or in astrophysics the emission line gas,
provides information about the physical processes of interest.  For
the astrophysical jets, the information one would like to obtain from
such analysis concerns two traditionally separated branches of
astrophysics:

The considerable energy release that may be associated with the jet
phenomenon is received by a large reservoir of gas surrounding the
host galaxy.  Compression and heating of that gas may influence the
star formation history of the host galaxy, in both ways, triggering
and suppression of star formation \citep{BLR96,Deyea97,Tortea09}.
Also, the jets transfer a bulk momentum to the emission line gas
surrounding the galaxy. The observed speeds in the emission line gas
are typically of order
the escape velocity of the host system \citep{Nesea08}.  This may lead
to observable effects in the spectral energy distribution of that
galaxy at later times \citep[e.g.][]{Tortea09}, and might even be part
of the answer of the question, why the mass of the supermassive black
holes correlate with properties of the host galaxies
\citep[e.g.][]{HR04}. 

The other branch is the physics of active galactic
nuclei. Extragalactic jets originate in the immediate vicinity of
super-massive black holes.  How the jet production is really happening
is not entirely clear yet, partly because the horizon scale, which is
the characteristic size of a black hole, cannot be resolved by any
instrument in any object. Also, an adequate treatment of the Kerr
metric, which permits angular momentum exchange in both ways between
the black hole and the plasma in the surrounding accretion disk
via magnetic torques, is
still a challenge to current magnetohydrodynamic (MHD) studies of accretion
physics, though progress is being made \citep{Hawlea07}.  Within the
theory of General Relativity, black holes have, in principal three
basic properties: mass, angular momentum and charge. As usual in
astrophysics, the charge is assumed to be negligible. Black hole
masses are in general well measured.  Because the horizon scale is
proportional to the mass, the mass, together with the accretion rate,
sets the scale for the overall energy output.  Jet production might be
coupled to the black hole's angular momentum and magnetosphere.  The
latter may be considerably different for different environmental
conditions, especially accretion rates. Some black holes develop a
strong quasar activity having at the same time powerful jets, while
some do not. Common sense suggests that, since masses and accretion
rates do not seem to make a difference, the black hole's angular
momentum, together with the magnetosphere, might be the decisive point
\citep[e.g.][]{BZ77,Cam91,McNea09}.  important evidence for or against
this hypothesis might in future come from correlating reliable
measurements of black hole spins and jet powers.  Black hole spins
have been measured from relativistically deformed iron lines of near
horizon material \citep[e.g.][and references
therein]{MC04,Youea05}. The values found include rapidly
spinning black holes close to the limit allowed by the Kerr metric
(cosmic censorship conjecture).  Jet powers have so far been
constrained from the interaction with the ambient gas, mainly from
X-ray data \citep[e.g.][]{Kr05,Birzea08}, but also first attempts have
been made to use the energy in the emission line gas for this
\citep{Nesea08}.

Jet beams are nearly dissipationless, and transport various kinds of
information to large scales, where they might in principle be deduced
from careful observation and analysis of their interaction with the
environment: the jet's power, mass flux, particle composition, net
electric current, and also the sense of the toroidal as well as the
direction of the axial magnetic field should be conserved along the
jet beam. Extended emission line regions may in principle serve as
calorimeters. Indeed, the emission line power is correlated with the
radio power \citep[e.g.][]{McC93}.  If shock ionisation is firmly deduced 
from the emission line ratios, we may conclude that the power source
of the radiation is the jet. Also, the kinetic energy in the emission line
gas, which can now be quite accurately determined from integral field 
spectroscopy is a lower bound for the energy released by the jet.
For two powerful radio galaxies at a redshift of about two, \citep{Nesea08}
give the observed radio power, power in the H$\alpha$ line and the
power required to account for the kinetic energy in the emission line gas:
Each value is a few times $10^{45}$~erg/s.

Here, we present hydrodynamics (HD) and magnetohydrodynamics simulations of the 
interaction of powerful radio jets with their environment with an emphasis 
on the cold gas ($T<10^6$K), which includes the emission line gas. We constrain
the locus of the emission line regions as well as the fraction of the jet power
they receive by global simulations in section~\ref{glob}.
We present local box simulations of the 
multi-phase turbulence on small scales in section~\ref{mpt}, which can be used to 
infer properties of the emission line gas like kinematics or condensation rate from 
the hot gas entrained into the radio cocoons.
We discuss our results and present the conclusions in section~\ref{last}

\section{Global jet simulations}\label{glob}
\begin{figure}
\centerline{\includegraphics[width=0.55\textwidth]{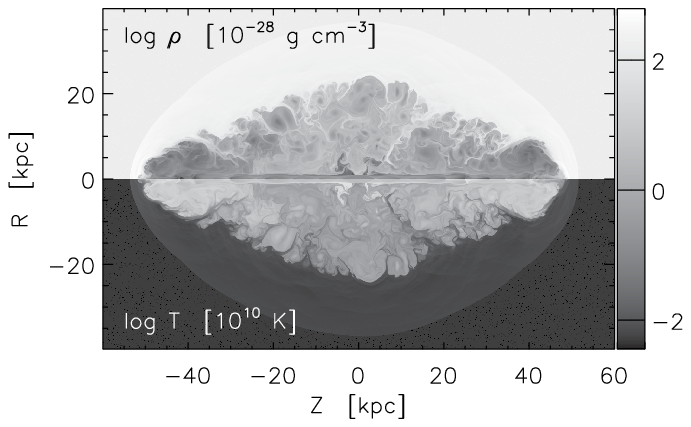}
  \includegraphics[width=0.45\textwidth]{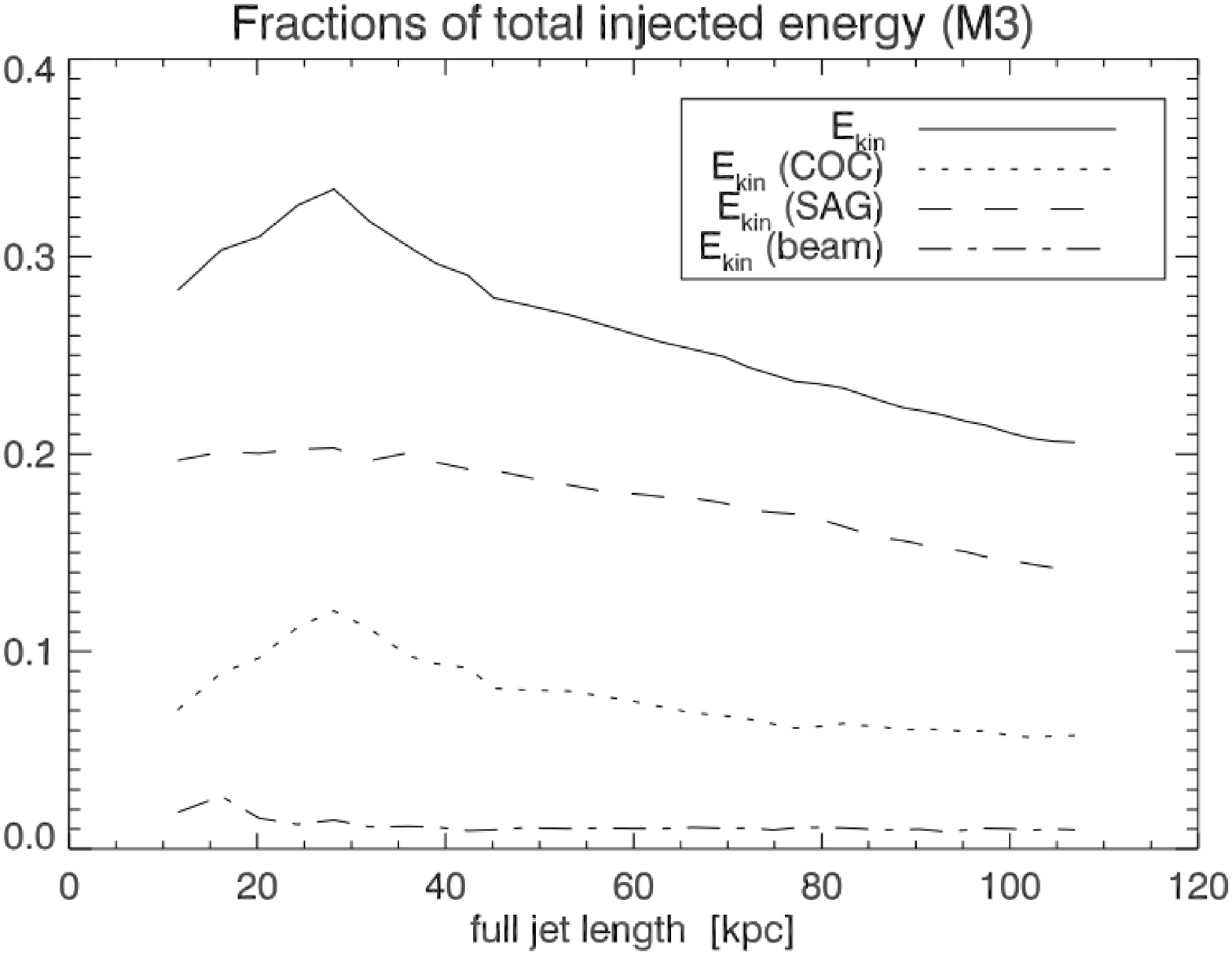}}
\caption{\small Left: MHD simulation of a very light (density ratio $10^{-3}$)
jet. Upper part shows the density, lower part the temperature. Right: Kinetic
energy as fraction of the total injected energy, for all the simulation and selected
regions explained in the text. The remaining energy is thermalised.}
\label{globjet}
\end{figure}

We have recently carried out a parameter study of global jet simulations with 
the MHD code \textit{NIRVANA} \citep{ZY97,GCK08,GKC09}.
The jets are injected with Mach~6 into a constant ambient density environment.
Here, we use the results for the simulation with a density ratio of $10^{-3}$.
We use an innovative magnetic field configuration: a helical magnetic field 
constrained to the jet material, which is almost zero in the ambient medium.
The global morphology of this simulations is shown in Fig.~\ref{globjet}.
The narrow jet beams propagate essentially undisturbed from the centre of the 
image out to the edges of the low density cocoon. The supersonic beams are shocked there
at the so-called Mach disk. The kinetic energy is transfered into thermal and produces
the hot spots there. Because of the large density contrast,
the excess pressure of the hot spots drives comparatively slow bow shocks into the
surrounding gas, which may be seen in Fig.~\ref{globjet} as an elliptical 
structure surrounding the source. Because of the slow bow shock expansion, 
the jet plasma has to flow backwards from the hot spots. At the same time it expands
sideways till it reaches pressure equilibrium with the surrounding layer 
of shocked ambient gas. The resulting width of the low density cocoon is anti-correlated 
to the jet/ambient density ratio. The large cocoon widths observed in extragalactic
radio jets usually require light jets, often the inferred jet density is 
three to four orders of magnitude below the ambient density.
The simulations have shown that very light jets produce a thick layer of shocked
ambient gas, and weak bow shocks. Where the bow shock can be identified in X-ray data,
both predictions are confirmed \citep[e.g.][]{Nulea05}.
The contact surface between the low density cocoon and the shocked ambient gas
layer, though partly stabilised by magnetic fields, is ragged by 
Kelvin-Helmholtz instabilities. Here, the ambient gas is entrained into the radio cocoon,
where it is accelerated by the turbulent cocoon motions.
Turbulence enhanced cooling as proposed below may enable such gas to cool down and 
contribute to the observed emission line halos.

\begin{figure}
\centerline{\includegraphics[width=0.5\textwidth]{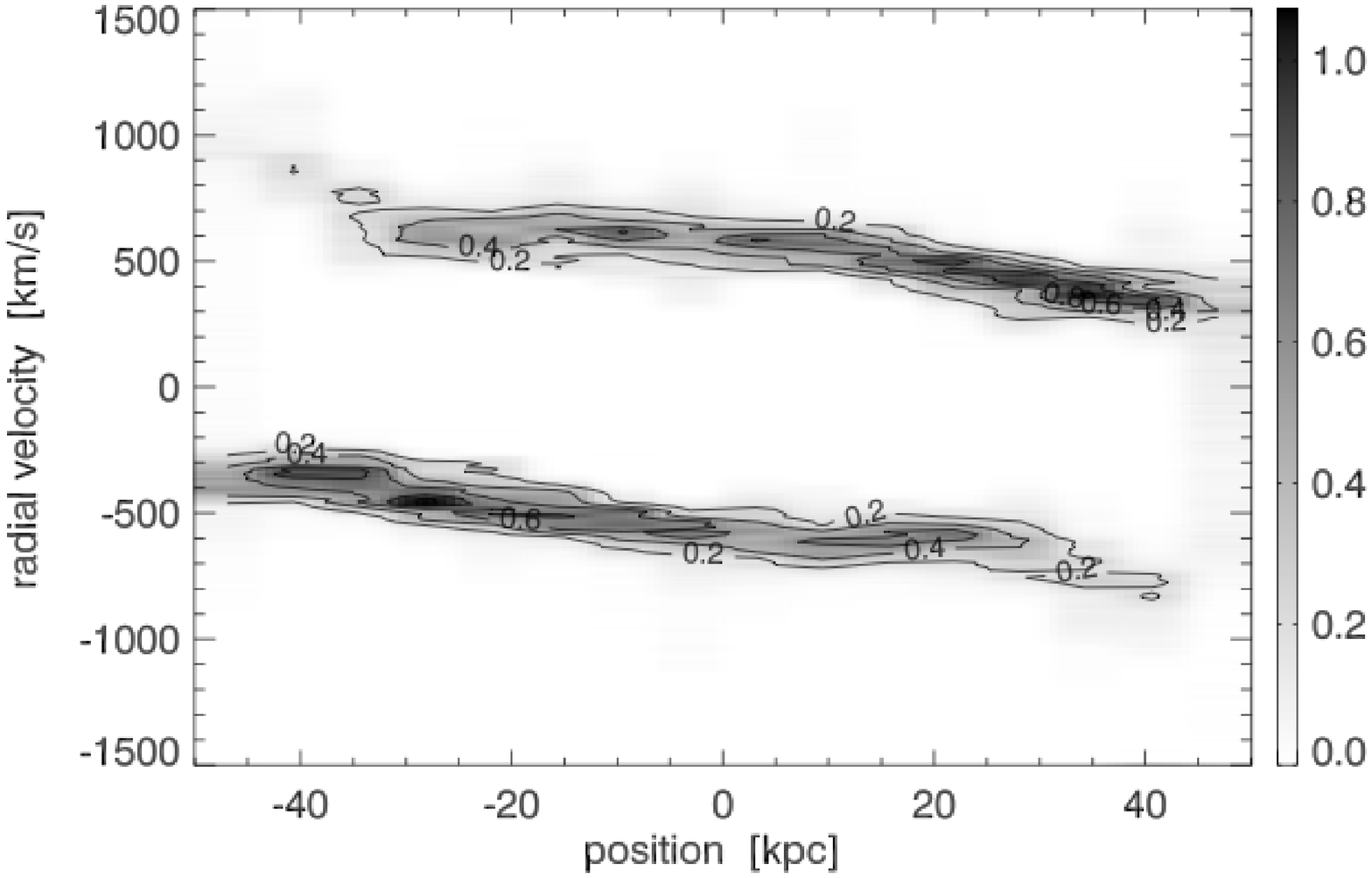}
  \includegraphics[width=0.5\textwidth]{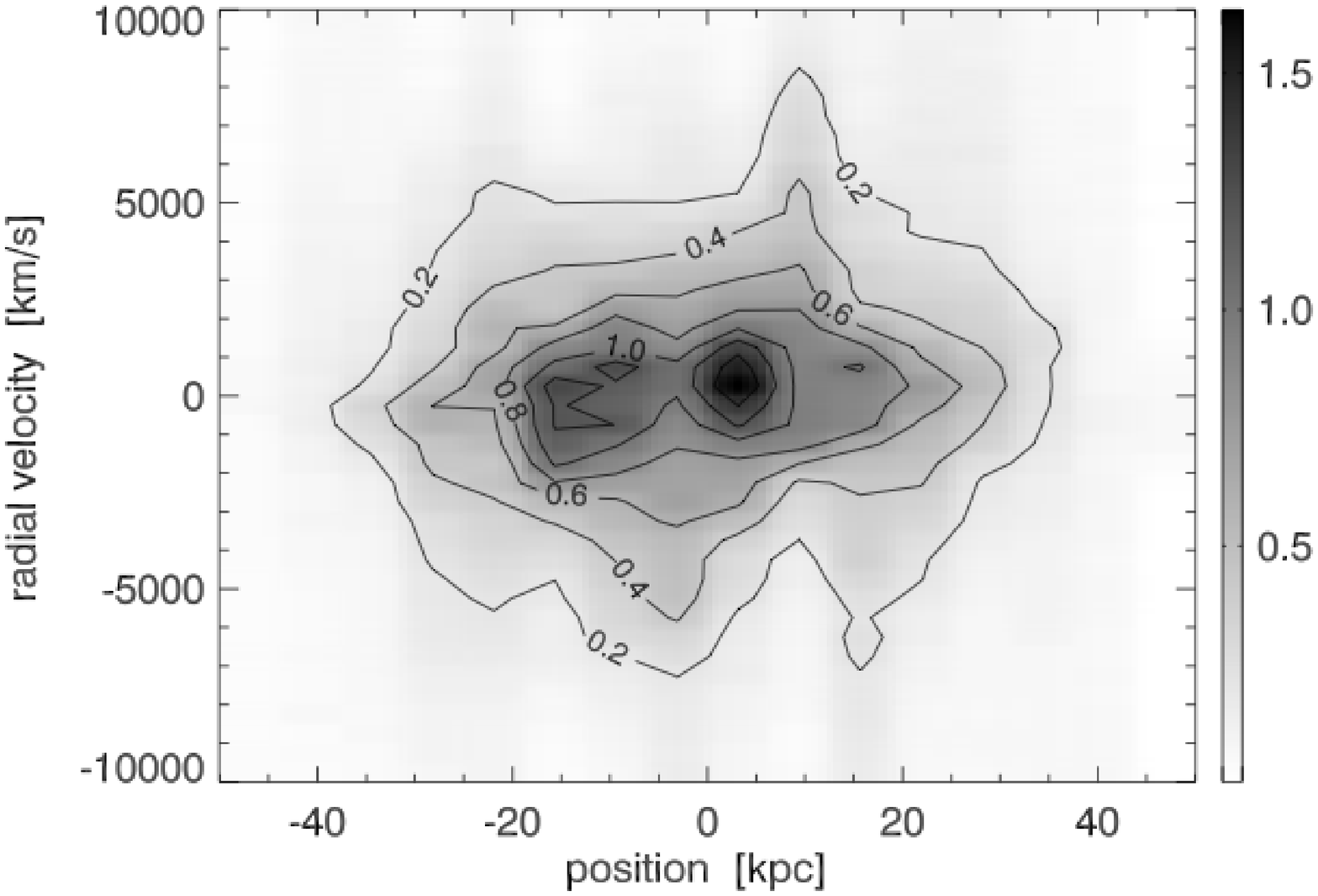}}
\caption{\small Position velocity diagram for the simulation of Fig.~\ref{globjet}.
Left: Cells in the shocked ambient gas. Right: Cells in the cocoon.}
\label{xv}
\end{figure}

An important question is, where the emission line gas is actually located with 
respect to the structures in the jet simulation. Many authors have assumed that it
should be the layer of shocked ambient gas, where the emission line gas actually 
resides. An alternative location would be the low density jet cocoon (cocoon hereafter).
From observations (compare Nesvadba, this volume) the emission line gas is usually
volume filling and does not show any kind of shell structure. The latter would 
be expected if the emission line gas would be located in the shocked ambient gas layer.
Therefore, the morphology strongly points to the cocoon as the emission line locus.
For an inclination of 70~degree from the jet axis,  we have calculated the position
velocity diagrams for the cocoon and shocked ambient gas layer, where the two have been
separated with a tracer technique (Fig.~\ref{xv}). The plot for the shocked ambient gas
displays the expected velocities, as they are similar to the sound speed in the 
ambient gas. However, the approaching and the receding part of the shocked ambient gas 
layer occupy clearly separated regions, which is usually not observed. 
A nice feature is the slope of the curve, which 
might be consistent with the observed large scale velocity gradients. 
The cocoon gas
occupies a coherent region. The velocities here refer to a numerical mixture 
of relativistic jet and hot entrained ambient plasma. Below, we will show, that
the multi-phase turbulence produces a characteristic density velocity relation, 
which can account for the observed velocities in the emission line gas. 
A remaining problem for the
cocoon locus are the bulk velocities: Cocoons in light jets flow backwards from the
hot spots, in the observers frame. Emission line regions are observationally inferred
to move outwards. We will address this problem further below.

We have calculated the kinetic energy available for cloud acceleration 
in the two respective regions in question (Fig.~\ref{globjet}, right). 
For our reference simulation, 
the kinetic energy in the cocoon is in the range 5-10~per cent of the total injected 
energy. For the shocked ambient gas layer, the number is 15-20~per cent.
These numbers are anti-correlated with the jet/ambient density ratio.
Within the accuracy of the measurements both situations would require a similar 
conversion between the measured kinetic energy in the emission line regions and the 
total jet power.


\section{Local simulations of multi-phase turbulence}\label{mpt}
\begin{figure*}[t!]
\resizebox{0.5\hsize}{!}{
\includegraphics[clip=true]{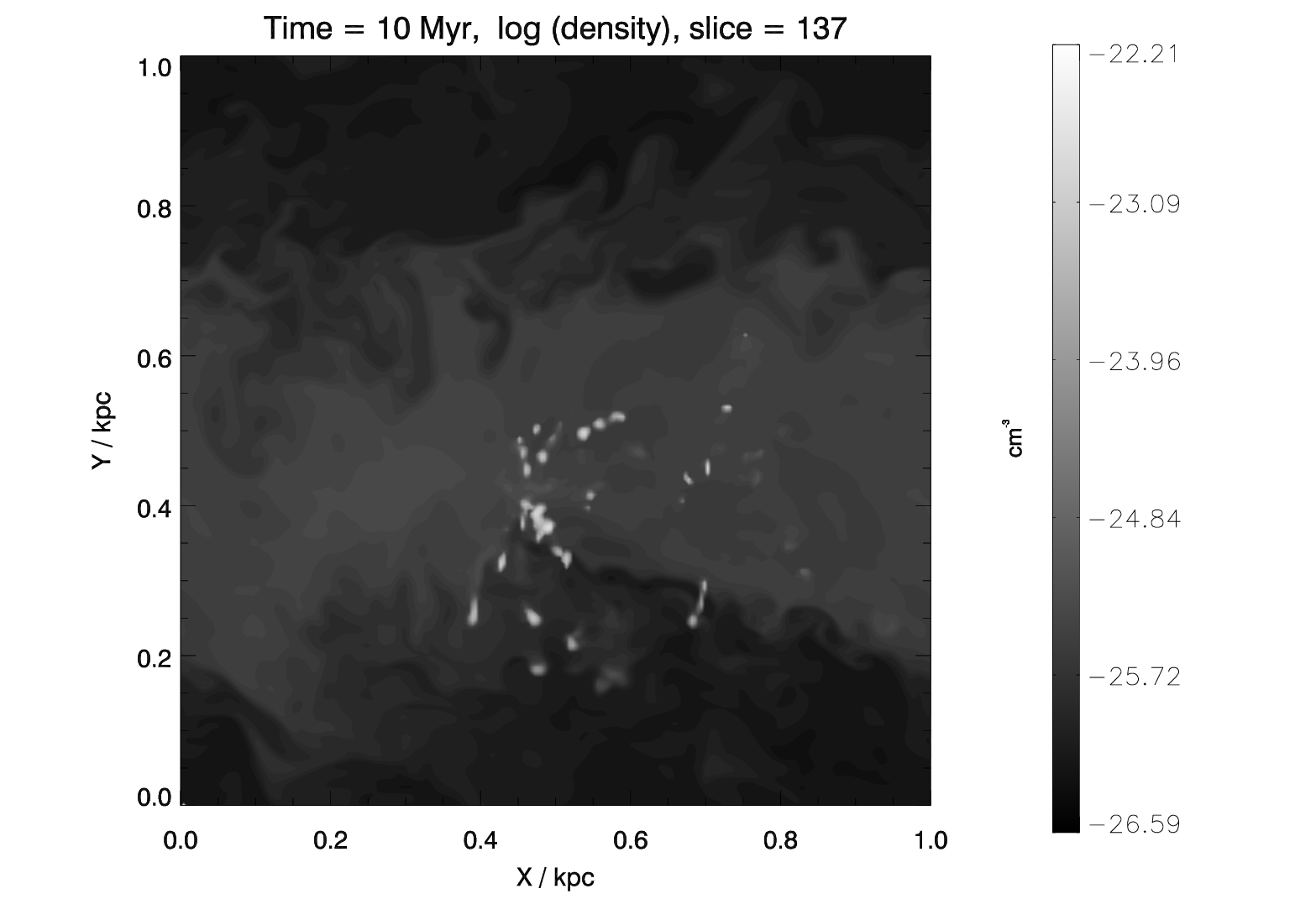}}
\resizebox{.47\hsize}{!}{
\includegraphics[clip=true]{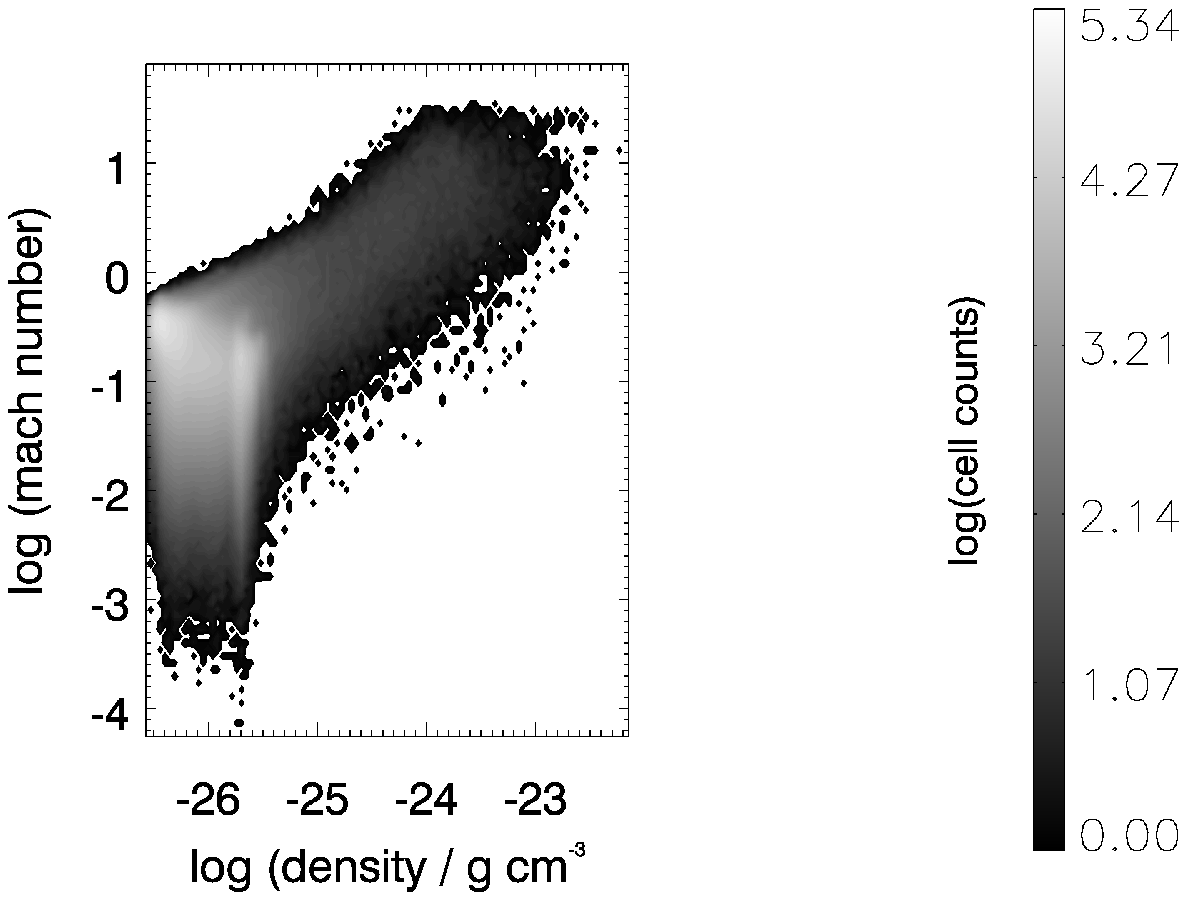}}
\caption{\footnotesize
Left: Density slice through the $\mu$1T5HR (see caption of Fig.~\ref{plots2}, 
for a description of the numbers) simulation at 10~Myr.
Right: Mach number density histogram for the same simulation. }
\label{plots0}
\end{figure*}

The actual emission line gas component cannot be resolved well in global 
jet simulations. We have therefore done local 3D HD simulations, including 
optically thin cooling, also with
\textit{NIRVANA}. We simulated essentially the Kelvin-Helmholtz instability
with a density ratio of $10,000$ and a dense cloud (1~and~10~cm$^{-3}$)
in the middle. The initial Mach number is 0.8 (80) in the hot (intermediate)
phase, respectively. In a series of simulations, we vary the cloud density
and the temperature of the intermediate (ambient) temperature phase.
The cloud is soon disrupted and the
non-linear evolution of the instability leads to multi-phase turbulence
(Fig.~\ref{plots0}, left). There are several numerical effects which may play a role
here. They are discussed along with more details about the simulations in
\citep{Kr08}. We believe however that the results we focus on here
are not dominated by the numerical issues.

\begin{figure*}[t!]
\resizebox{.45\hsize}{!}{
\includegraphics[clip=true]{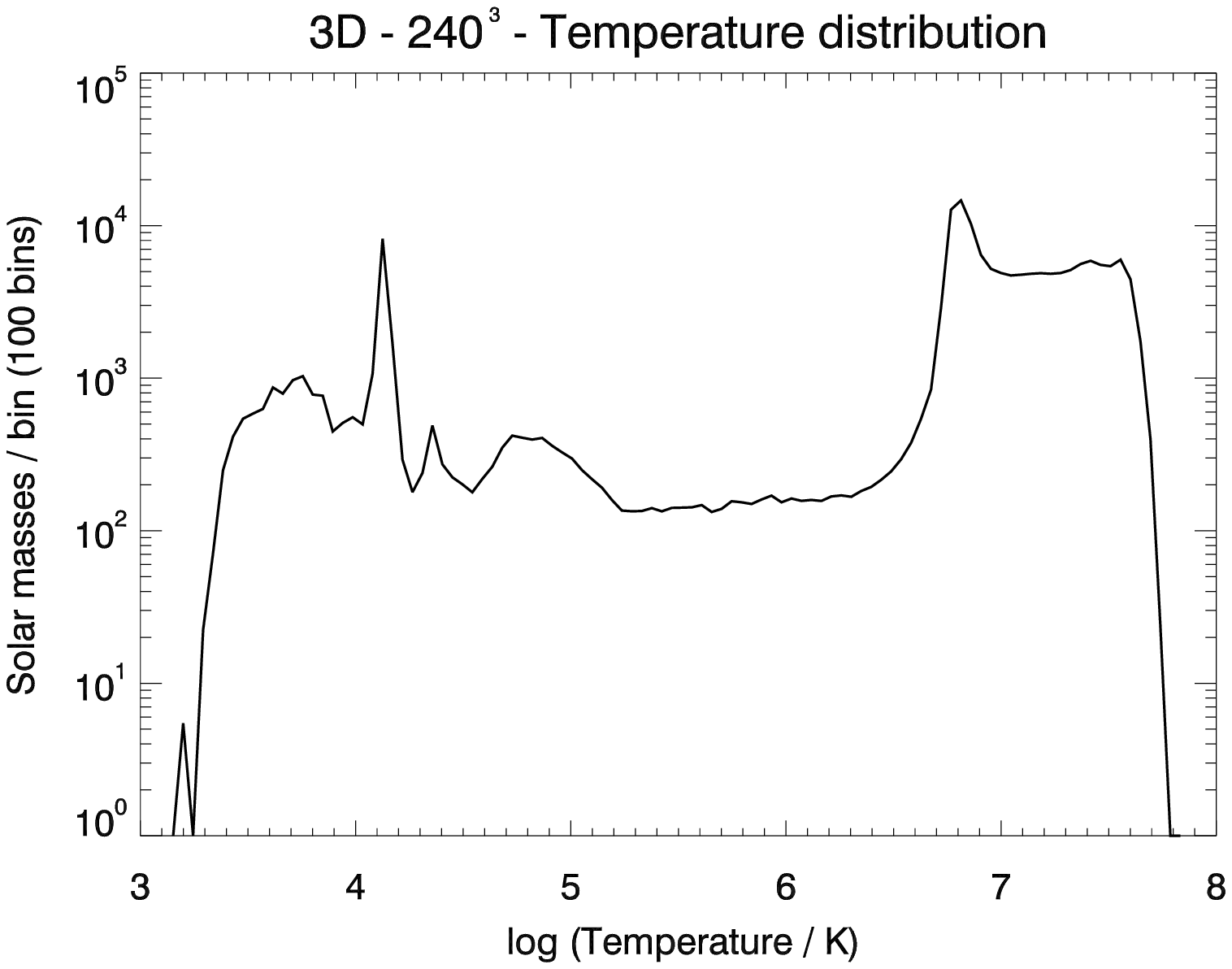}}
\resizebox{.45\hsize}{!}{
\includegraphics[clip=true]{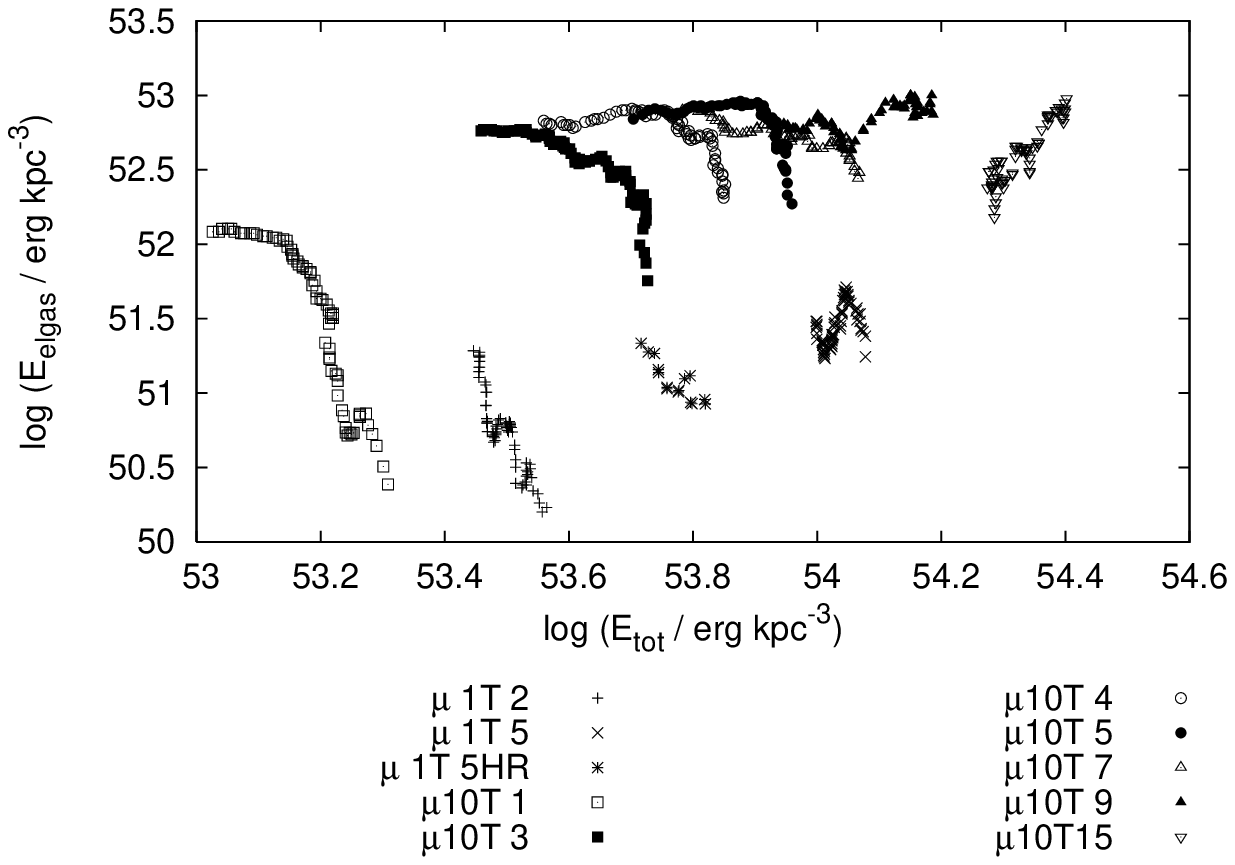}}
\caption{\footnotesize
Left: Temperature histogram ($\mu$1T5HR). Shock excitation and cooling together
produce the pronounced peak at $10^4$~K, here identified with the emission line gas. 
Right: Energy in the emission line gas versus total energy in the simulation.
The simulations with low mass load have a lower fraction of their energy in the 
emission line gas. The time evolution is from high total energy to lower total energy.}
\label{plots1}
\end{figure*}

We show a typical temperature histogram in Fig.~\ref{plots1} (left). 
The intermediate temperature gas is still well visible at $5 \times 10^6$~K.
Numerical mixing produces the flat regions at both sides of this peak.
Towards the lower temperature end the cutoff is given by the resolution limit
of the simulation. Near $10^4$~K, a prominent peak is found. This may be interpreted 
as shock excited emission line gas. In the following analysis we define 
gas with a temperature between 10,000~K and 20,000~K as emission line gas. 

Fig.~\ref{plots0} (right) shows the Mach number - density histogram. The Mach number
is not correlated  to the density at low densities, where the cooling time is much
longer than the simulation time. When cooling is important, the Mach number is
proportional to the square root of the density. The result is identical to the 2D
case \citep{KA07}. It implies that the velocity distribution of all gas colder 
than $\approx 10^6$~K should be the same. Also, from the critical density when cooling
becomes important down to the density of jet material, the velocity scales
with the square root of the density. The velocity of the backflow of the jet plasma
in the cocoon should still be of order the speed of light. The critical density
is probably not very much higher than the ambient density in these strongly 
star forming galaxies. The observed width of the emission line gas then requires 
jet/ambient density ratios of order $10^{-4}-10^{-3}$, similar to what can be deduced
from X-ray observations of sources at low redshift \citep{Kr03,Kr05}.

We show the energy fraction in the emission line gas in Fig.~\ref{plots1} (right).
Due to the radiative dissipation, the time evolution for each of the different 
simulations is from high to lower total energy. Except for the highest ambient 
temperature run, the energy in the emission line gas rises first, and for many
simulations levels off at a constant energy. For those simulations,
the energy in the emission line gas is a few ten per cent of the total energy.
This fraction depends strongly on the mass load, with all the low mass load simulations
having a much lower energy fraction in the emission line gas. 
Our simulations have a mass load of $\approx 10^5$M$_\odot$~kpc$^{-1}$, roughly
three orders of magnitude less than observed extended emission line regions.
All this suggests that the latter have most of their turbulent energy in the 
emission line gas. There could also be a significant fraction in a yet colder
component, which our simulations do not resolve well enough.

\begin{figure*}[t!]
\resizebox{.47\hsize}{!}{
\includegraphics[clip=true]{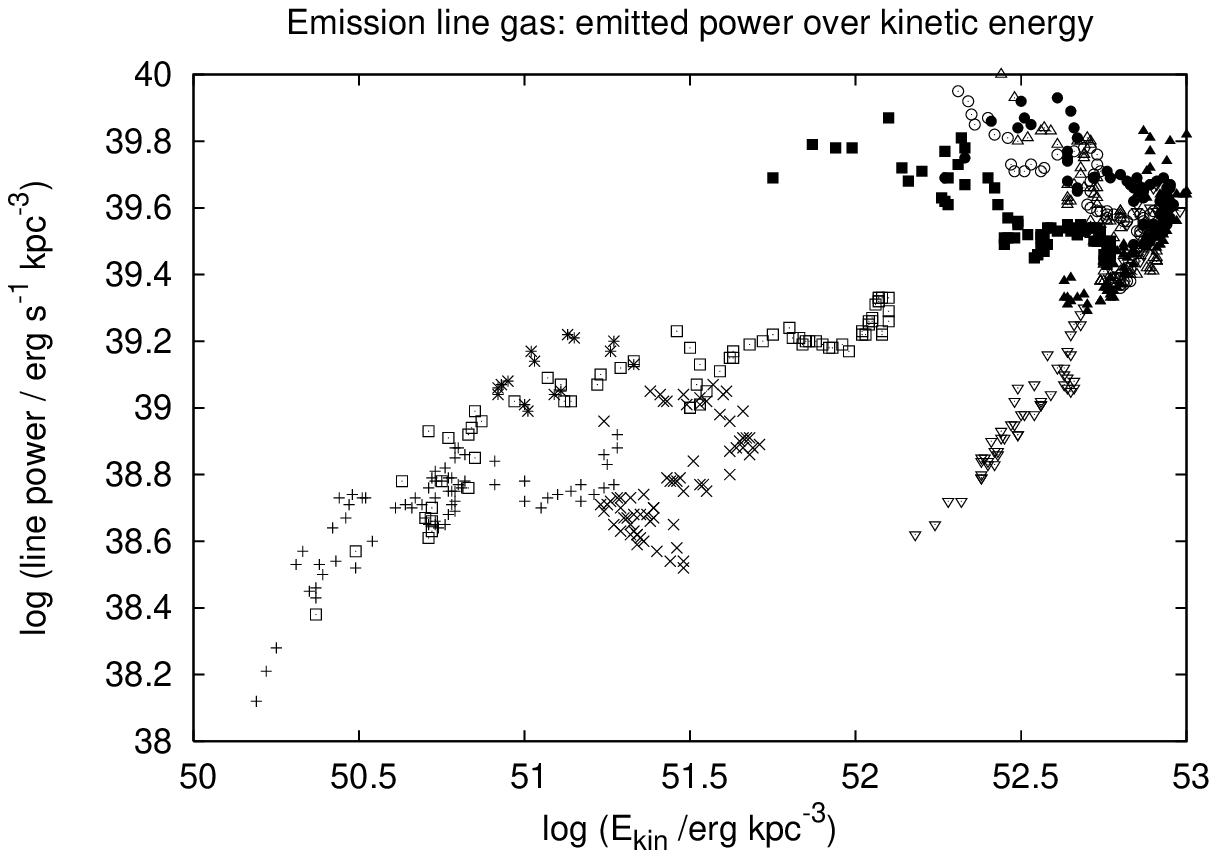}}
\resizebox{.47\hsize}{!}{
\includegraphics[clip=true]{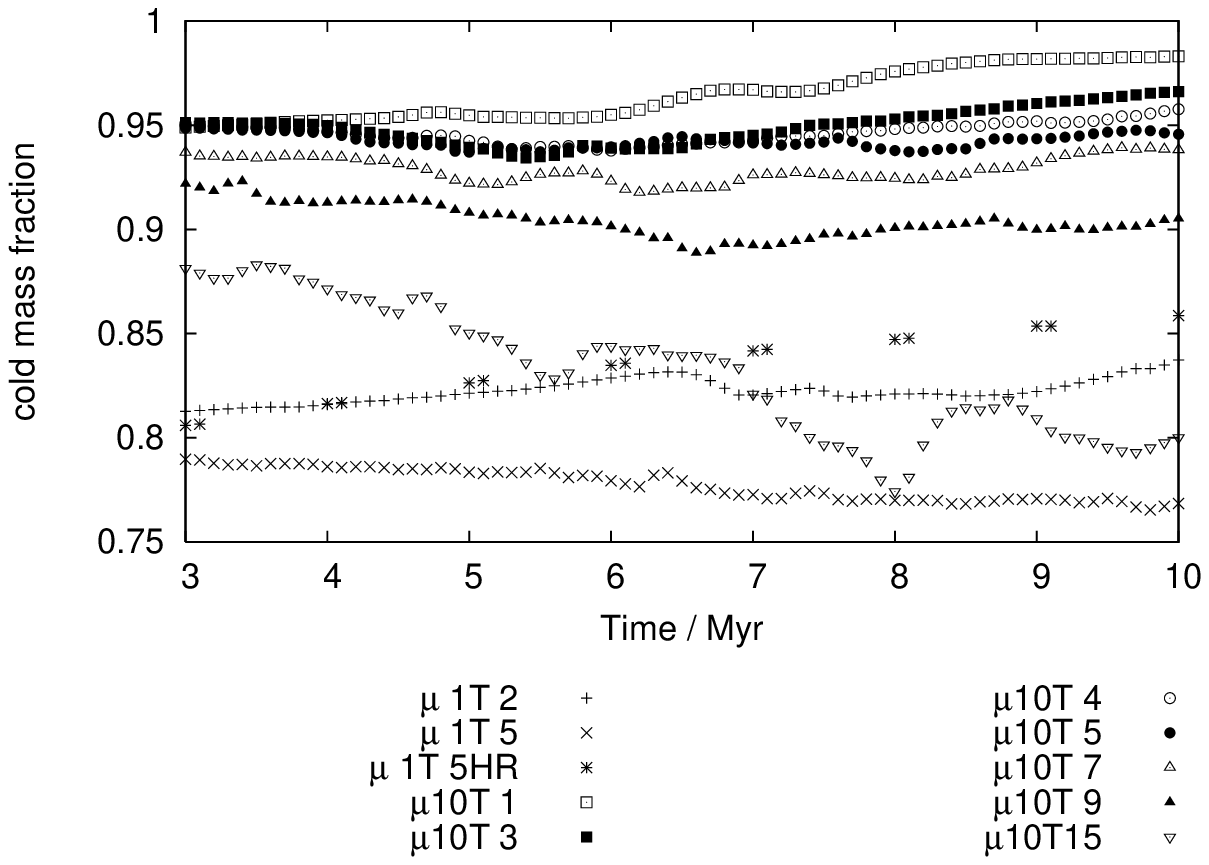}}
\caption{\footnotesize
Left: Radiative power, emitted by the emission line gas (temperature in the
range from 10,000~K to 20,000~K) over Kinetic energy of this
gas. Different symbols are used for different runs shown at many times, 
as indicated in the legend.
$\mu$xTy denotes an initial cloud density of x~m$_\mathrm{p}$~cm$^{-3}$
and an initial temperature of the intermediate component of y~Mio.~K.
Right: Cold gas ($T<10^6$~K) fraction over time. }
\label{plots2}
\end{figure*}

Our simulations also suggest a relation between the kinetic energy in the 
emission line gas (far more than the thermal energy in this gas phase) and the 
radiated power (Fig.~\ref{plots2}, right). At low kinetic energy,
about $10^{12}$~erg are needed to radiate at 1~erg/s, or equivalently,
the radiative dissipation timescale, which is the relevant dissipation
timescale for multi-phase turbulence, is $10^{12}$~s. At higher energy,
the dissipation timescale has a wider spread for the different simulations
with a tendency to increase to about $10^{13}$~s. For the observed 
emission line regions in high redshift radio galaxies, this number
is of order $10^{14-15}$~s, consistent with the observation that the
turbulence has just decayed in the larger sources. Again, the kinetic 
energies we reach in our simulations fall short of the observationally
required energies by three to four orders of magnitude, which
might well explain the discrepancy in the dissipation timescales.

Finally, we show the time evolution of the cold ($T<10^6$~K) gas fraction
in (Fig.~\ref{plots2}, right). Depending on the temperature of the intermediate
gas, which also determines the total energy in the box, the cold gas fraction
is either increasing or decreasing. For about 5 Mio~K the initial cold gas 
fraction of about 95 per cent seems to be right. The result depends on the 
numerical resolution in the way that we underestimate the cold gas fraction.
In summary, for ambient temperatures of up to 5~Mio~K, in equilibrium,
most of the gas mass should be in the cold phase.
From these results, it seems plausible in principle that much of the 
emission line gas is cooled from entrained warmer ambient gas, with the
help of some initial cold gas filaments.

\section{Discussion and conclusions}\label{last}
We have presented global jet simulations, as well as local box simulations 
of multi-phase turbulence. Regarding the locus of the emission line gas in 
high redshift radio galaxies, from the global simulations, both 
the shocked ambient gas and the radio cocoon have problems. The most severe
for the shocked ambient gas is the expected shell morphology and the split
nature of the position velocity diagram. While dust may play some role in 
individual sources, the rarity of these features in observed sources
argue against this scenario. The cocoon model requires a detailed model
for the multi-phase turbulence in the cocoon. We give such a model
with our box simulations, and show that kinematics and radiated
power can be explained. We even can give a possible origin for the 
emission line gas, namely turbulence enhanced cooling on a small seed
cold gas fraction. A remaining problem are the bulk velocities:
While the simulations predict inward motion during the active jet phase,
the observations, based on the Laing-Garrington-effect, show outflowing
gas. Perhaps, a way to make sense of this would be to assume that much of the 
emission line gas is actually not gas entrained from the halo 
into the cocoon in a warm phase, but rather initially kinematically cold
gas within the galaxy. If the turbulent radio cocoon would interact with
a massive gas disk, as those observed at similar redshifts by Foerster Schreiber et al.
\citep{FSea09}, one would expect a turbulent diffusion into the radio lobes,
with about the observed bulk flow properties. 
This conclusion is supported by a consideration of the momentum 
in the emission line gas and the jet, respectively.
For one bubble of emission line gas the momentum is:
\begin{equation}
P_\mathrm{elg}=10^{51} \,\mathrm{g\, cm \, s^{-1}}
 M_\mathrm{elg,10}\,  v_\mathrm{elg,500}\, ,
\end{equation}
where $M_\mathrm{elg,10}$ is the mass of the emission line gas in units
of $10^{10} M_\odot$ and $v_\mathrm{elg,500}$ is the bulk outflow velocity
in units of 500~km~s$^{-1}$
\citep{Nesea08}. If Cygnus~A would be regarded as a prototype for 
FR~II radio sources, then the total mass that has been driven through the 
jet beam for a $\approx 100$~kpc sized source would be of order:
\begin{equation}
M_\mathrm{jet}= 10^4 M_\odot \dot{M}_{-3}\, t_7 \, ,
\end{equation}
where $\dot{M}_{-3}$ is the jets mass flux in units of $10^{-3} M_\odot$~yr$^{-1}$,
and $t_7$ is the age of the source in units of 10~Myrs.
The total momentum delivered by the jets is hence typically of order 
\begin{equation}
P_\mathrm{jet}= 10^{48} \,\mathrm{g\, cm \, s^{-1}}
 M_\mathrm{jet,4} \Gamma_3 \, .
\end{equation}
Again, the mass gone through the jet beam is given in convenient units of 
$10^4 M_\odot$ and the jets bulk Lorentz factor was scaled to three.
So, unless the jets are extremely relativistic, which seems unlikely,
the jet's momentum falls short of the one required to accelerate the emission
line gas by orders of magnitude. So, while the jet energy is enough
to cause turbulent motion, the momentum is too small to be responsible
for the directed motions.
Further support for this scenario comes from the HI~shells surrounding
small high redshift radio galaxies, only \citep{Kr05b}. They have a typical diameter
of 50~kpc, and a typical velocity of 200-300~km/s. The energy required
to drive these shells can therefore be precisely calculated and 
is of order one supernova per year, corresponding to star formation rates 
in excess of 100~$M_\odot$~yr$^{-1}$. This is actually found in observations
\citep{FSea09}.
If this is hence indeed evidence for the finishing of a starburst,
one could interpret the emission line halos as the missing observational 
evidence for the late phase in galaxy merger simulations with active black holes
\citep{DMSH05}: Due to the shape of the cavities drilled by the jets, the expelled
gas is not distributed isotropically, but aligned with the radio jets.

What happens to the gas, when the active jet phase terminates?
We have carried out simulations of the long term effect of radio sources
representing the emission line gas by tracer particles \citep{HKA07}.
Powerful jets set up a long term gas convection that may persist for many Gyrs.
From these simulations, we predict the emission line gas to be distributed
in the ambient gas, contributing to the halo metallicity. 

Focusing again on the black hole, the emission line regions give power
estimates close to the emitted radio power. Together they approach $10^{46}$~erg/s.
The jets must have considerably more than this to keep propagating.
We showed above that overall of order 1 per cent, maybe up to 10 per cent
of the total jet energy
should be expected to appear as kinetic energy in the emission line gas
Hence, extended emission line regions point to kinetic jet powers of order
$10^{47}$~erg/s or greater. This is of order the Eddington power for a $10^9 M_\odot$
black hole.
If the accretion state of the central quasar could be sufficiently constrained,
one might be able to infer, whether the jet is powered by accretion directly
or by the spinning down of the central black hole.


\end{document}